\documentclass{ws-rv975x65}
\usepackage{subfigure}     
\usepackage{ws-rv-thm}     
\usepackage{ws-rv-van}     
\usepackage{url}
\usepackage{unites2e}

\makeindex

\begin{document}

\chapter[Atmospheric Cherenkov Gamma-ray Telescopes]{Atmospheric Cherenkov
  Gamma-ray Telescopes}\label{ACTs}

\author[J. Holder]{Jamie Holder}

\address{Department of Physics and Astronomy and the Bartol Research Institute,\\
University of Delaware, Newark DE19716,\\
jholder@physics.udel.edu}

\begin{abstract}

  The stereoscopic imaging atmospheric Cherenkov technique, developed
  in the 1980s and 1990s, is now used by a number of existing and
  planned gamma-ray observatories around the world. It provides the
  most sensitive view of the very high energy gamma-ray sky (above
  $30\U{GeV}$), coupled with relatively good angular and spectral
  resolution over a wide field-of-view. This Chapter summarizes the
  details of the technique, including descriptions of the telescope
  optical systems and cameras, as well as the most common approaches to data
  analysis and gamma-ray reconstruction.


\end{abstract}

\body

\section{Introduction}

\begin{figure}
\centerline{\includegraphics[width=\textwidth]{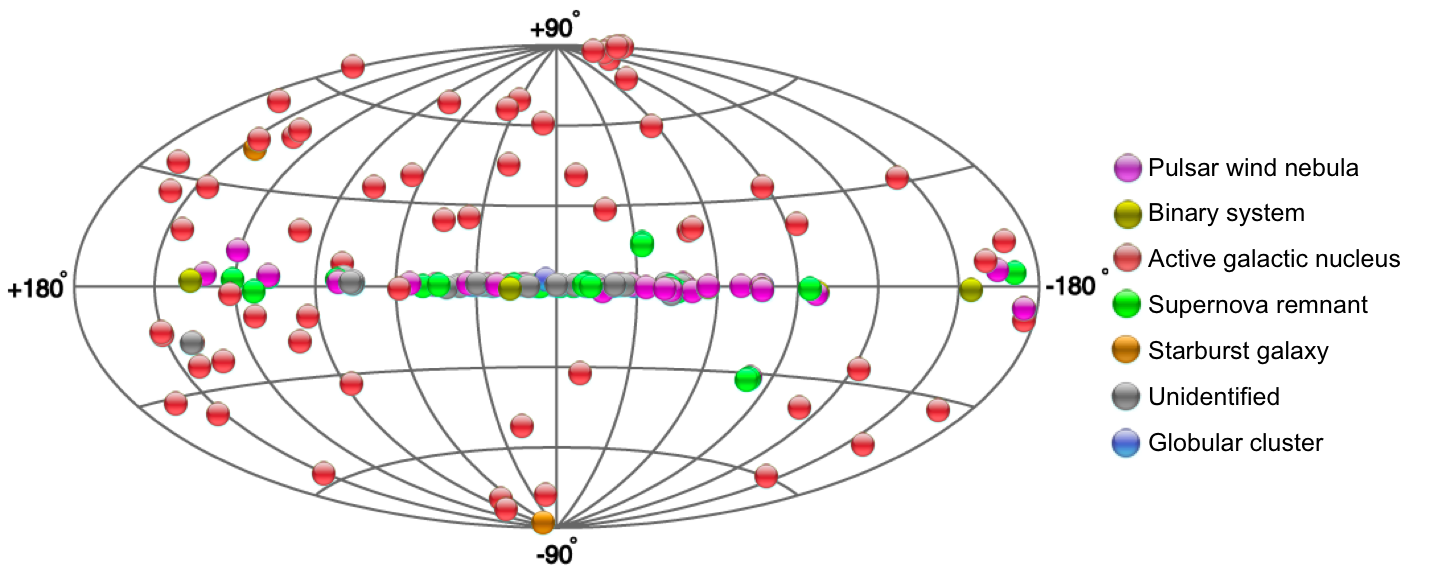}}
\caption{The locations, in Galactic coordinates, of all known
  astrophysical sources of TeV gamma-ray emission, as of
  2015 (Figure courtesy of TeVCat.\cite{TeVCat} See there for a
  full description of the different source classes). 
 }
\label{TeVsky}
\end{figure}

Astrophysical very high energy (VHE) gamma-rays (with energies
$\gtrsim30\U{GeV}$) are believed to result almost exclusively from the
interactions of populations of highly relativistic particles with
ambient matter or photon fields. The study of these VHE photons
therefore allows us to examine the processes of particle acceleration
in the Universe, and the extreme environments in which they
occur. Gamma-ray astronomy also provides a unique tool for many
complementary astrophysical topics. For example, extragalactic
background photon fields and intergalactic magnetic fields can be
measured, or constrained, by their imprint on the measured properties
of distant gamma-ray sources. Gamma-ray signatures of candidate dark
matter particles may also lie in the VHE band, and can be sought
through observations of regions in which the densest clumps of dark
matter are believed to lie. Around 150 VHE gamma-ray sources have now
been detected\cite{TeVCat, 2012APh....39...61H}
(Figure~\ref{TeVsky}). These comprise many different source classes
(pulsars and their nebulae, supernova remnants, and active galactic
nuclei, to name a few) and the majority have been discovered using
ground-based Atmospheric Cherenkov Telescopes (ACTs).

The Earth's atmosphere is opaque to high energy photons, and so the
most direct approach to study the gamma-ray sky is to send detectors
into space. However, astrophysical gamma-ray production mechanisms
typically result in steeply falling power-law spectra, leading to a
very low photon flux at high energies. The Crab Nebula, for example -
one of the brightest astrophysical gamma-ray sources - produces a flux
of only $\sim6\U{photons}\UU{m}{-2}\UU{year}{-1}$ at the Earth above
$1\U{TeV}$. To study the Universe at these energies therefore requires
a detector with enormous collection area, far beyond the maximum
practical size of a satellite-borne device (which is
$\sim1\UU{m}{2}$). Atmospheric Cherenkov telescopes achieve this feat
by measuring the Cherenkov light produced by gamma-ray triggered
particle cascades (or \textit{air showers}) in the atmosphere. In this
way, using the Earth's atmosphere as an intrisic part of the detection
technique, effective collection areas can easily exceed $10^5\UU{m}{2}$.

The potential of this approach for gamma-ray astronomy was first
explored by Jelley and Galbraith in the
1950s\cite{1955JATP....6..304J}, but attempts to exploit it were
hampered by the overwhelming background of charged cosmic rays. The
first significant discovery of an astrophysical TeV gamma-ray source
was not made until the detection of the Crab Nebula, using the Whipple
10-meter telescope, in 1989\cite{1989ApJ...342..379W}. This success
was the result of the development of effective methods to record an
\textit{image} of the Cherenkov emission from air showers. A complete account
of the long history and development of the field is given by Hillas.\cite{2013APh....43...19H}

\begin{figure}
\centerline{\includegraphics[width=\textwidth]{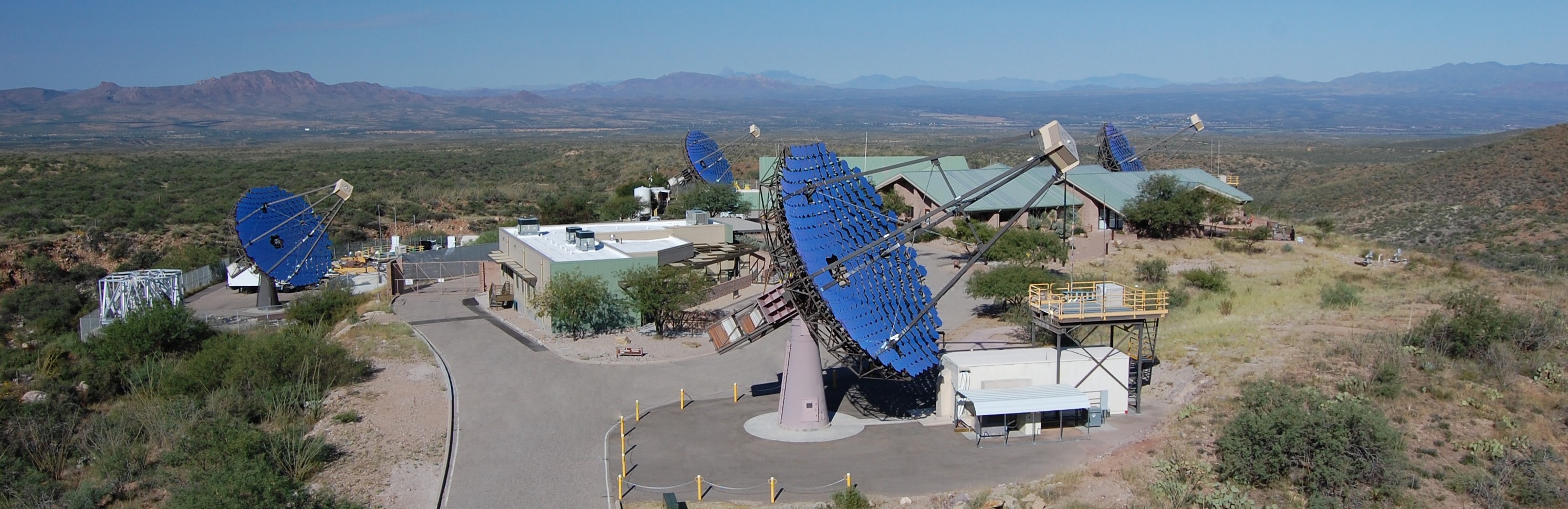}}
\caption{The VERITAS atmospheric Cherenkov Telescope array.\cite{veritas} 
 }
\label{VERITAS}
\end{figure}

Three major ACT facilities are currently operating, the key properties of
which are listed in Table~\ref{arrays}. They each provide sensitivity
to gamma-ray sources with a flux below $1\%$ of the steady flux from
the Crab Nebula. Figure~\ref{VERITAS} shows one of these, the VERITAS
array, with which the author is associated.

\begin{table}[ht]
\tbl{Details of each of the major atmospheric Cherenkov telescope facilities.}
{\begin{tabular}{@{}lcccccc@{}} \toprule
  & Location & Number & Aperture & Optical & Number      & Field- \\
  &                & of telescopes &       & design & of pixels   & of-view \\
\colrule
H.E.S.S.                   & Namibia        & 4 & $12\U{m}$ & Davies-Cotton & 960 & $5.0^{\circ}$ \\
H.E.S.S. II$^{\text a}$& Namibia        & 1 & $28\U{m}$ & Parabolic & 2048 & $3.2^{\circ}$ \\
MAGIC II                 & La Palma        & 2 & $17\U{m}$ & Parabolic & 1039  & $3.5^{\circ}$ \\
 VERITAS                  & Arizona, USA & 4 & $12\U{m}$ & Davies-Cotton & 499  & $3.5^{\circ}$ \\
\botrule
\end{tabular}
}
\begin{tabnote}
$^{\text a}$H.E.S.S. II is an addition to the H.E.S.S. array, located in
the centre of the four original telescopes.
\end{tabnote}
\label{arrays}
\end{table}

\section{Air Showers and Atmospheric Cherenkov Emission}

The design of atmospheric Cherenkov gamma-ray telescopes is driven by
the essential characteristics of Cherenkov emission from air showers,
which we first briefly describe.

A VHE gamma-ray incident on the Earth's atmosphere converts into an
electron-positron pair. Subsequent Bremsstrahlung and pair production
interactions lead to the generation of an electromagnetic cascade in
the atmosphere. The radiation length, $X_0$, for Bremsstrahlung in the
atmosphere is $37.15\U{g}\UU{cm}{-2}$, which is 7/9 of the mean free
path for pair production.  This similarity allows a simple analytical
approximation for the shower development (first developed by Heitler\cite{1954qtr..book.....H}), in which the total number of electrons,
positrons and photons doubles every $\mathrm{ln}(2)X_0$. The
primary gamma-ray energy, $E_0$, is split evenly among the secondary
products. The shower continues to develop until the average electron
energy drops to $E_c=84\U{MeV}$, the critical energy below which
ionization losses dominate. The maximum number of particles in the
cascade is given by $E_0/E_c$.

Cosmic rays - charged, relativistic protons and nuclei - also produce
air showers in the atmosphere. In this case, the
cascade development is more complex, with hadronic interactions
proceeding through a variety of channels, leading to the production of
secondary nucleons, along with charged and neutral pions with large
transverse momenta. The pions do not survive to sea level: neutral
pions decay rapidly into two gamma-rays, while charged pions produce
muons and neutrinos:
\[
\pi^o \longrightarrow \gamma + \gamma
\]
\[
\pi^+ \longrightarrow \mu^+ + \nu_{\mu}
\]
\[
\pi^- \longrightarrow \mu^- + \bar{\nu}_{\mu}
\]

\begin{figure}
\centerline{\includegraphics[width=10cm]{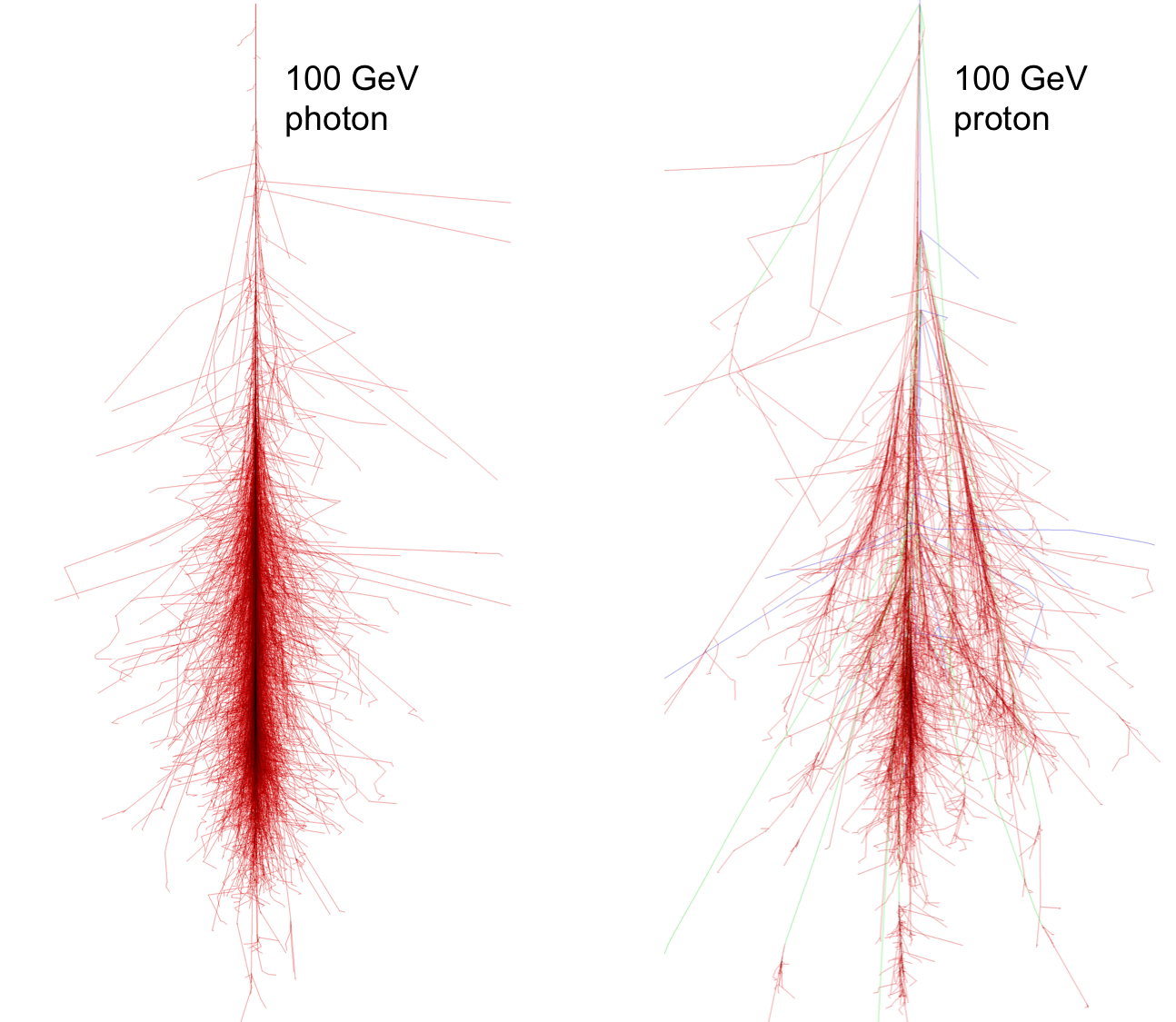}}
\caption{ Monte Carlo simulations of the tracks of particles in photon
  and proton initiated air showers.\cite{CORSIKApics} The first interaction height is
  fixed at $30\U{km}$. The horizontal axis range is $\pm5\U{km}$.
 }
\label{showers}
\end{figure}

The gamma-ray secondaries thus produced can trigger electromagnetic
sub-showers, while the long-lived muons form the most penetrating
component of the cascade, often reaching the ground. The result of
this is that cosmic ray initiated air showers develop much less
regularly than gamma-ray initiated cascades, as illustrated in
Figure~\ref{showers}. These differences in the shower morphology,
along with the reconstruction of the arrival direction of the incoming
primary, allow ACTs to achieve an efficient discrimination of
gamma-ray photons from the otherwise overwhelming isotropic cosmic ray
background.

\begin{figure}
\centerline{\includegraphics[width=\textwidth]{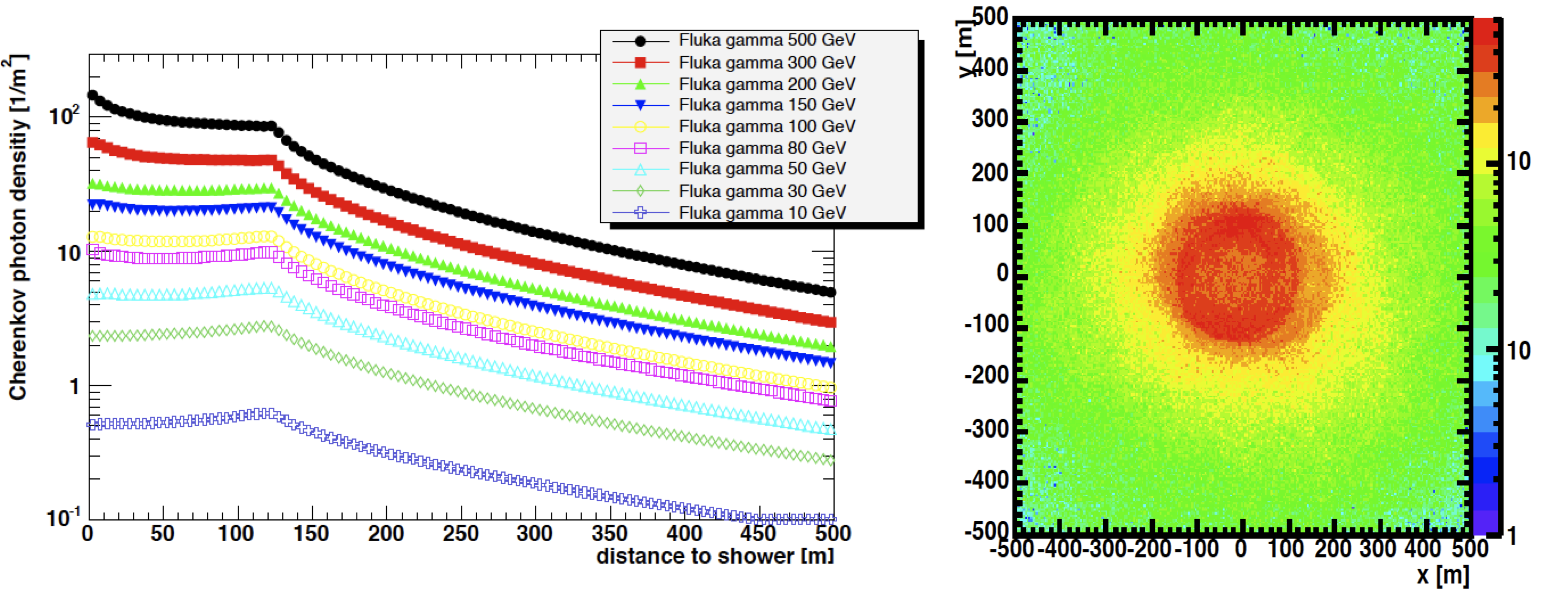}}
\caption{ Monte Carlo simulations of the distribution of Cherenkov
  photons on the ground for gamma-ray initiated air showers. The left
  plot shows the Cherenkov photon density as a function of radial
  distance from the shower core for primaries with a range of
  energies, the right shows the two-dimensional photon density on
  the ground for a shower with a $300\U{GeV}$ primary. Figure courtesy
of G. Maier.}
\label{lightpool}
\end{figure}

The relativistic charged particles in air showers are moving faster
than the speed of light in air ($v>c/n_{air}$, where $n_{air}$ is the
refractive index) and so generate Cherenkov radiation. Cherenkov light
is produced throughout the cascade development, with the maximum
emission occurring when the number of particles in the cascade is
largest, at an altitude of $\sim10\U{km}$ for primary gamma-ray
energies of $100\U{GeV}$ to $1\U{TeV}$. Each particle generates
Cherenkov light at a fixed angle to the direction of
motion, ($\theta_C$), given by
\[
\mathrm{cos}\theta_C=\frac{c}{vn_{air}}
\]
The Cherenkov angle is $\sim1.3^{\circ}$ at sea level. Electromagnetic
cascade particles also undergo multiple Coulomb scattering, which
distributes their directions over a small angular range and generates
the shower's lateral extent. The resultant filled ``pool'' of Cherenkov
light on the ground has a photon density of
$\sim100\U{photons}\UU{m}{-2}$ for a $1\U{TeV}$ gamma-ray primary, and
a radial extent with a peak at $\sim130\U{m}$, as illustrated in
Figure~\ref{lightpool}. The peak is due to a focussing effect
resulting from the changing angle of Cherenkov emission with
atmospheric depth.

The Cherenkov photon yield is proportional to $1/\lambda^2$ (where
$\lambda$ is the wavelength). The spectrum is therefore dominated by
blue/UV emission, peaking around 340\U{nm}. Shorter wavelength
emission is subject to atmospheric absorption (particularly ozone),
and therefore does not reach the ground, unless it is generated very
deep in the atmosphere (for example by penetrating muons).  Cherenkov
photons from each shower arrive in a brief pulse of a few nanoseconds
duration. The time-averaged photon yield from all air showers
constitutes only $\sim1/10000$th of the background night sky light in
the visible, but the light from a single shower can rival the
brightest objects in the night sky for the brief duration of the
pulse.

\section{Detection}
The goal of an atmospheric Cherenkov gamma-ray telescope is to detect
the Cherenkov emission from air showers, and to use this to determine
the nature of the primary (gamma-ray or cosmic ray), along with its
arrival direction and energy. The detection technique is, in essence,
rather simple, requiring only a large mirror to collect Cherenkov
photons, and a fast photon detector coupled to an oscilloscope to
record them. The first detection of Cherenkov emission from an air
shower was made with a $0.2\UU{m}{2}$ reflector, a single
photomultiplier tube (PMT) and a free-running analog oscilloscope.\cite{1953Natur.171..349G} Modern ACT arrays perform the same
task, but can reach mirror areas $>600\UU{m}{2}$, instrumented with
thousands of PMTs coupled to GHz sampling electronics and
sophisticated trigger systems.

While detection is relatively straightforward, gamma-ray
discrimination and reconstruction is rather more challenging. One
approach is to measure the arrival time and photon density
distribution of the Cherenkov light at ground level. This ``wavefront
sampling'' method was explored by experiments such as STACEE\cite{2005ITNS...52.2977G} and CELESTE\cite{2002NIMPA.490...71P},
using the very large mirror areas provided by the heliostats of
existing solar power facilities. The brightest previously known
astrophysical gamma-ray sources were detected using this technique,
but the difficulty of effective gamma-ray discrimination
limited its usefulness. The technique may be more applicable at the
highest energies($>10\U{TeV}$), where small, widely separated
detectors allow to achieve effective areas of $\sim100\UU{km}{2}$. This
idea is currently being investigated by the HiSCORE experiment.
\cite{2014APh....56...42T}

\begin{figure}
\centerline{\includegraphics[width=0.7\textwidth]{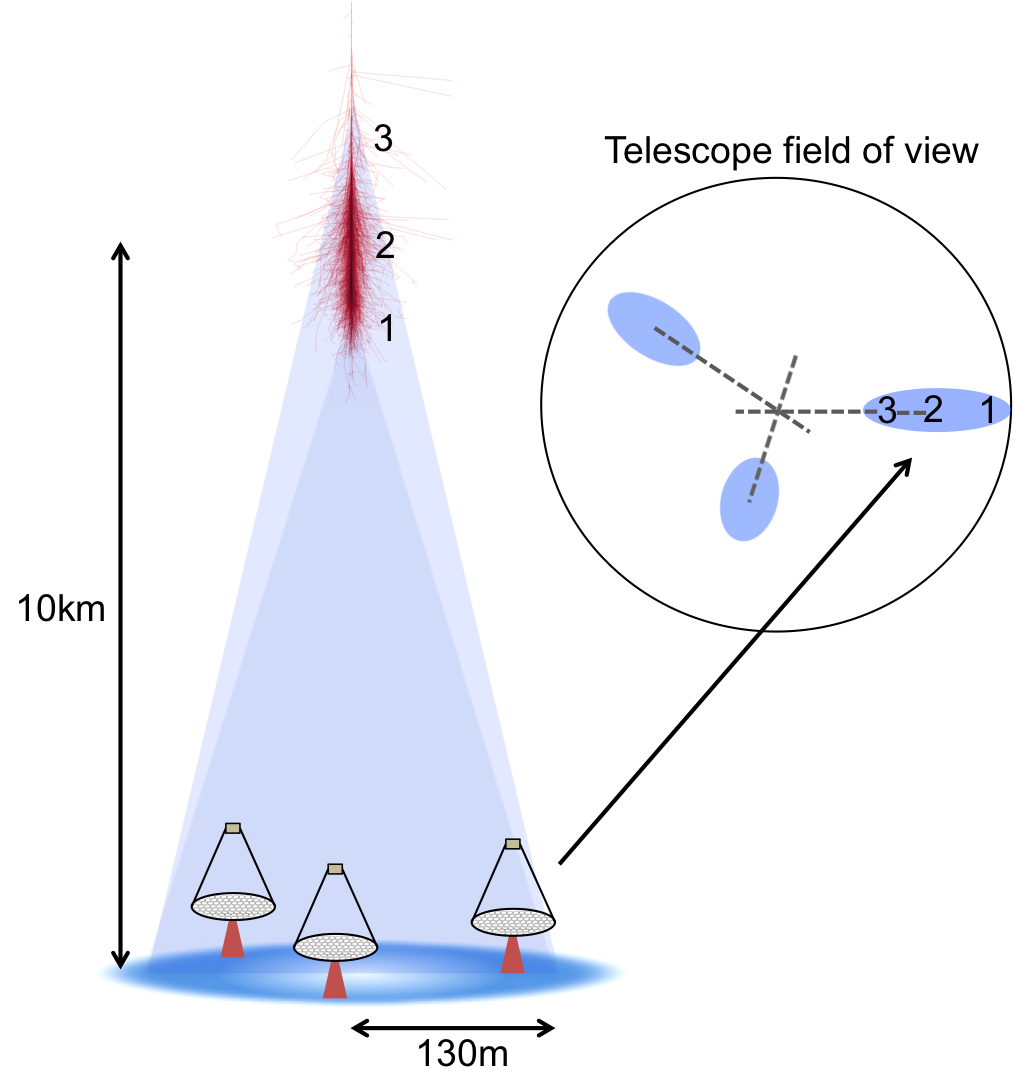}}
\caption{ An illustration of the stereoscopic imaging technique. A
  gamma-ray triggers an electromagnetic cascade in the Earth's
  atmosphere, which generates Cherenkov radiation in a pool on the
  ground. Telescopes within this light pool are used to form an image
  of the shower, which allows reconstruction of the arrival direction of
  the incident primary photon.}
\label{schematic}
\end{figure}

By far the most successful approach, used by all of the major
facilities in operation today, is the stereoscopic imaging
technique. The principle of this is illustrated in
Figure~\ref{schematic}. Large convex reflectors are used to focus the
Cherenkov light from air showers onto a camera comprised of
photo-detector pixels. The camera records an image of the shower, and
the properties of the image (its shape, intensity and orientation),
allow determination of the properties of the shower primary. Applying this
to an array of telescopes (``stereoscopy'') provides a view of the
same shower from a number of different perspectives, and so enhances
the geometrical shower reconstruction. It is worth stressing that a
key aspect of this technique is the necessity for accurate Monte Carlo
simulations of both the shower development and the detector response.

\section{The Design of an Atmospheric Cherenkov Telescope Array}

Since they observe blue Cherenkov light from air showers, atmospheric
Cherenkov gamma-ray telescopes are effectively optical telescopes,
working in the visible band of the electromagnetic spectrum. They are
subject to the same constraints as other optical telescopes -
observations must be conducted at night, under clear skies, at a dark
site - but the design requirements are very different. 

\subsection{Optical systems}

Two competing requirements inform the optical design of ACTs. The
first is for a very large aperture, and hence mirror area. This allows
one to collect as many Cherenkov photons as possible from each shower,
which in turn defines the lowest gamma-ray energy threshold of the
telescope. Fortunately, the relatively crude cameras of ACTs, and the
lack of detailed structure in the Cherenkov images,
means that the mirror form and surface quality is much less important
than for optical telescopes. An optical point-spread-function of
a few arcminutes is usually adequate. This level of performance can
be achieved using tessellated reflectors, made up of hundreds of
individual mirror facets.

The second requirement is for a large field-of-view. Cherenkov images
from air showers are approximately elliptical in shape, with an angular
extent of up to a few degrees. The images are offset from the arrival
direction of the shower primary - in the case of gamma-ray initiated
showers, this means that the image is offset from the gamma-ray source
position in the field-of-view.  The angular distance of the offset is
proportional to the shower impact parameter\footnote{The distance
  between the shower core projected onto the ground and the
  telescope.} (Figure~\ref{schematic}). Even a point source of
gamma-rays, therefore, requires a field-of-view of a few degrees
diameter. In reality, many known sources of gamma-ray emission
(particularly supernova remnants and pulsar wind nebulae) have a large
angular extent. Additionally, analysis of ACT data typically uses a
portion of the field-of view in which there are no known gamma-ray
sources to estimate the background of remaining cosmic ray
showers. Currently operating arrays have fields-of-view of
$3^{\circ}-5^{\circ}$, while plans for the next generation of
instruments reach $8^{\circ}-10^{\circ}$.

\begin{figure}
\centerline{\includegraphics[width=0.8\textwidth]{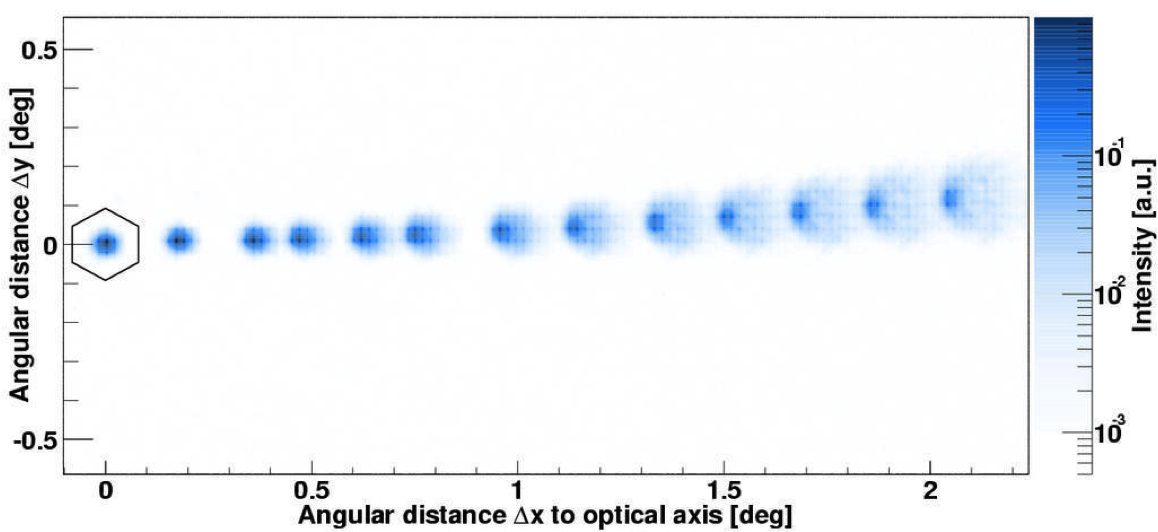}}
\caption{The image of a star viewed at different distances from the
  optic axis of a H.E.S.S. telescope, showing the effects of optical
  aberration in a Davies-Cotton optical system.\cite{2003APh....20..129C}}
\label{aberration}
\end{figure}

The requirement for a very large field-of-view for each telescope
dictates a small focal ratio (focal length, $f$, divided by aperture,
$D$) - typically around 1.0. Off-axis optical aberrations,
particularly coma and astigmatism, are therefore an important
consideration. A common approach for a tessellated reflector, used
extensively for ACTs (starting with the Whipple 10-meter), was first
developed by Davies and Cotton for a U.S. Army solar furnace facility
- their original application was for the thermal testing of materials
for military purposes.\cite{1957SoEn....1...16D} In this design,
individual spherical mirror facets, with a radius of curvature of
twice the focal length of the telescope ($2f$), are placed on the
surface of a spherical reflector with a radius equal to $f$. The
facets are aligned such that the normals of the individual facets
point to the $2f$ position along the optic axis. The reflector is
therefore discontinuous at every point, and ideal performance is
achieved with the smallest facets. As well as providing off-axis
performance superior to that of a single spherical or parabolic
reflector (Figure~\ref{aberration}), the Davies-Cotton design uses
identical mirror facets, which can be inexpensively
mass-produced. Mirror facet alignment is also relatively simple. One
downside, however,  is that the design is not isochronous - the reflector induces
some time spread in the arrival time of Cherenkov photons at the
telescope cameras, typically on the order of a few
nanoseconds. Tessellated parabolic reflectors, used by the world's
largest ACTs ( MAGIC and H.E.S.S. II), do not suffer from this
drawback, but require facets of varying forms to be produced, with a
corresponding increase in cost and complexity.

Aplanatic two-mirror telescopes provide a solution to off-axis
abberations, whilst retaining isochronicity and also reducing the plate
scale in the focal plane significantly.\cite{2007APh....28...10V} Cost and complexity again present
challenges, but the benefits of two-mirror systems are such that they
will very likely form a part of the next generation ACT arrays.
Prototyping is already underway, with the Schwarzschild-Couder design
among the favoured options.\cite{2008ICRC....3.1445V}

The technology for producing mirror facets is also a very active area
of development.\cite{2013SPIE.8861E..03P} Traditional techniques use
milled aluminium, or glass which is ``slumped'' to the required shape,
polished, and then coated with anodized aluminium. Carbon or glass
fibre, aluminium honeycomb or a composite design can offer a more
lightweight, cost-effective solution.  With typically hundreds of
mirror facets per telescope, alignment of the facets is not
trivial. Stepper motors can be used to provide active mirror control,
which greatly simplifies this task, as well as allowing for alignment
corrections due to mechanical deformations during observations.

\subsection{Telescope structure}

\begin{figure}
\centerline{\includegraphics[width=0.9\textwidth]{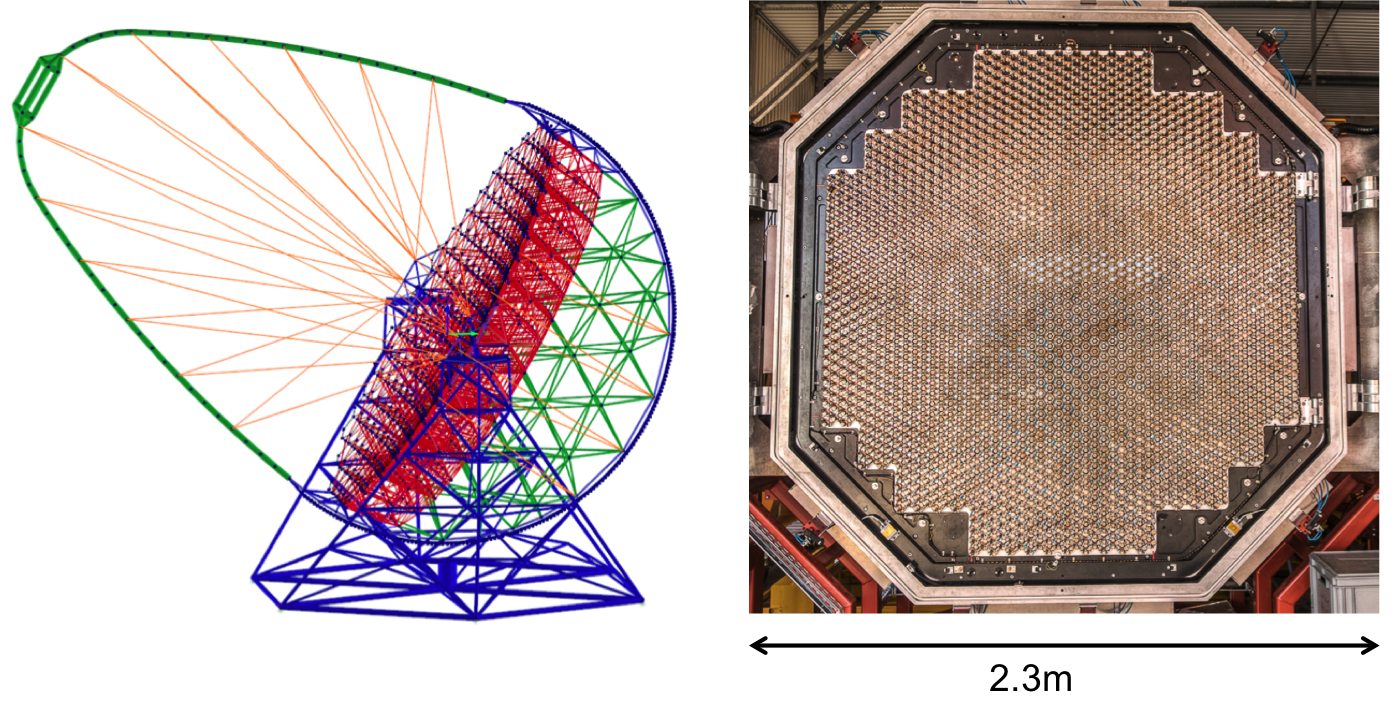}}
\caption{{\bf Left:} The carbon fibre support structure design for the
  $23\U{m}$ aperture Large Size Telescopes of the Cherenkov Telescope
  Array.\cite{2013arXiv1307.4565A} \newline {\bf Right:} The PMT camera of the
  H.E.S.S. II telescope, containing 2048 PMTs and weighing 3
  tonnes.\cite{2014NIMPA.761...46B}}
\label{scope_n_cam}
\end{figure}

The mechanical design of ACTs is also challenging, given the extremely
large apertures, and the necessity of supporting a large, delicate and
massive detector package at the prime focus. Weight and simplicity
considerations have led to the adoption of altazimuth mounts for all
modern ACTs. The rigidity requirements of the optical system and
camera support structures have been solved in two ways - either by the
brute force approach of constructing the telescope superstructure from
a steel space frame (in the case of the H.E.S.S. and VERITAS
telescopes, for example), or by the use of a lightweight carbon fibre
frame coupled with an active mirror adjustment system (in the case of
MAGIC, and planned for the largest next generation telescopes - see
Figure~\ref{scope_n_cam}).

The telescope is also required to track accurately - typically to
within a few arcminutes. The position of the telescope is monitored by
encoders (usually optical, with arcsecond resolution). A software
model of the telescope pointing, calibrated using observations of
stars, is used to translate these measurements into a position on the
sky. The online tracking is supported by CCD pointing monitors, which
are fixed to the telescope structure and track star positions, as well
as the exact gamma-ray camera location. Offline corrections based on
these CCD measurements are used to reduce systematic telescope
pointing errors to typically tens of arcseconds.

ACTs must also be able to slew to new targets as rapidly as
possible. The $40\U{tonne}$, $17\U{m}$ diameter MAGIC telescopes, for
example, are able to move to observe any position in the sky within 40
seconds. This requirement is driven by the transient nature of the
gamma-ray sky, which contains many sources known to flare dramatically
on short timescales. In the case of gamma-ray bursts, the emission may
last just a few seconds - although none of these have yet been
detected from the ground, despite rapid slewing triggered by satellite
alerts.

\subsection{Cameras}

Large aperture, single-reflector ACTs require physically large cameras
($\geq1\U{m}$) to cover an adequately large field-of-view. In order to
record an image, the photosensitive area must be divided into pixels,
numbering hundreds or thousands, with each pixel sampling
$\lesssim0.1^{\circ}$. The photodetector pixel of choice for ACTs has,
in most cases, been the photomultiplier tube (PMT). These devices
provide reasonable photon detection efficiency ($\sim20\%$),
nanosecond response times, a large detection area, and extremely clean
signal amplification (by factors of $\sim100,000$), allowing them to easily
resolve single photon signals. Dead space between the photo-sensitive
areas of the individual pixels is recovered by placing close-packed
light-concentrating Winston cones on the camera face. One example is
the H.E.S.S. II PMT camera, shown in Figure~\ref{scope_n_cam},
weighing 3 tonnes and containing 2048, $1.25\U{inch}$ PMTs. The
camera is housed at the telescope focal point, $36\U{m}$ from the
centre of the reflector dish, giving a field-of-view of $3.2^{\circ}$
in diameter (which is relatively small for an
ACT).\cite{2014NIMPA.761...46B} While the size of ACTs prohibit the
construction of domes around the complete telescopes, the expensive
and delicate PMT cameras are usually housed in light-tight boxes,
which allow for daytime testing and calibration. The H.E.S.S. II
camera can actually be removed when required, and stored
in a protective enclosure.

A number of recent technological advances are now finding their way
into ACT camera design. PMT photocathode developments now yield
quantum efficiencies of up to $40\%$ at short wavelengths (so-called
``ultra-bialkali'' devices). PMTs are now also available in
``multianode'' packages, in which an array of close-packed PMT cells
are incorporated in a single housing, greatly reducing the cost per
pixel of an ACT camera, and allowing for much finer pixellation of the
field-of-view. Silicon photodetectors also show great promise, as
demonstrated by the FACT (``First G-APD Cherenkov Telescope'')
telescope, a small ($9.5\UU{m}{2}$) pathfinder experiment, equipped
with a camera containing 1440 individual Geiger-mode avalanche
photodiode detectors.\cite{2011NIMPA.639...58A} Modern silicon-based
devices can provide higher photon detection efficiency than PMTs,
require lower operating voltages, and can cover large areas at
relatively low cost - in particular with the development of Multi-Pixel
Photon Counter (MPPC) arrays, containing arrays of up to 64 discrete
detectors, each of which can be read out individually.

 \subsection{Trigger and Data Acquisition Systems}

 The arrival time of atmospheric Cherenkov flashes at the telescope is
 random and unpredictable. The flashes also last just a few
 nanoseconds, and exhibit temporal structure on timescales even shorter
 than this. Continuously monitoring the sky with GHz sampling rates on
 hundreds or thousands of channels is impractical; instead, it is
 necessary to \textit{trigger} the data acquisition system of ACTs,
 such that the photo-detector outputs are recorded only for a small
 time window around the arrival time of the Cherenkov flash. Since the
 trigger decision time is longer than the duration of the flash
 itself, the photodetector output signals must be delayed (e.g. by
 routing analog signals through long cables), or continuously sampled
 and stored in digital memory buffers. Upon receipt of a valid
 trigger, the relevant data time window can be accessed and saved to
 disk as digital samples.

\begin{figure}
\centerline{\includegraphics[width=1.0\textwidth]{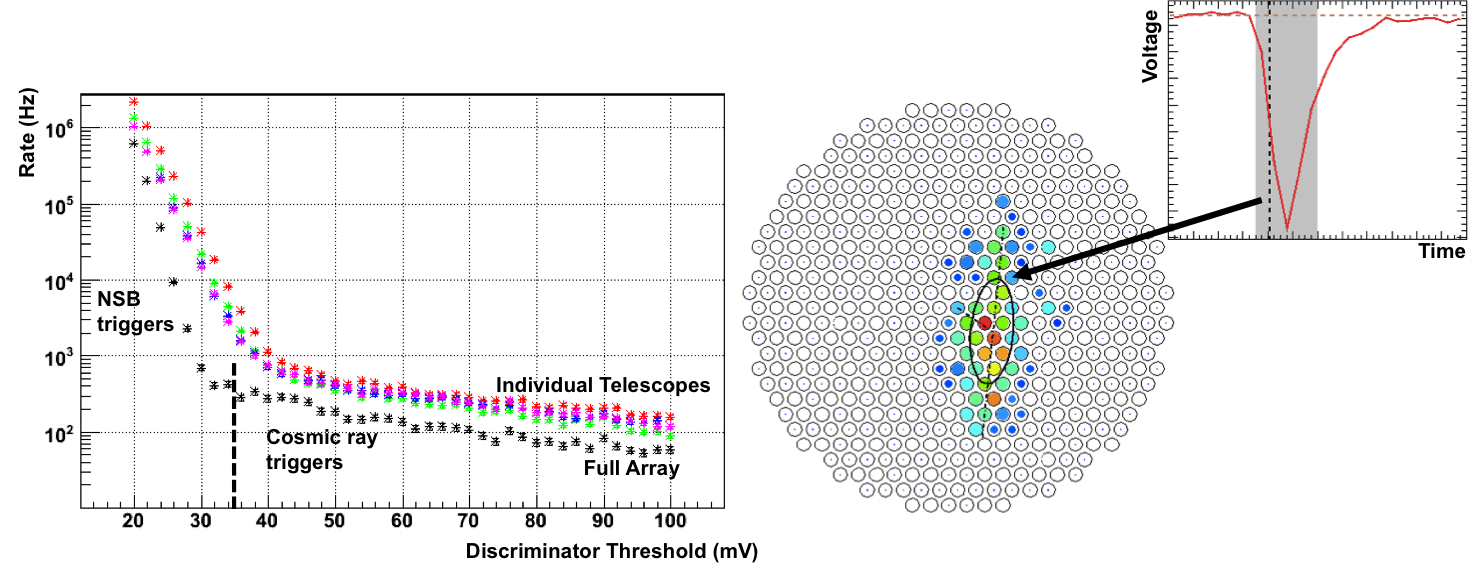}}
\caption{{\bf Left:} A``bias curve'' for the VERITAS array. The rate of triggers
  is shown as a function of the discriminator threshold chosen for all
  of the PMT pixels. The upper curves correspond to the individual
  telescope rates, the lower curve to the complete array, after
  requiring a coincident trigger from at least two telescopes. A clear
  break point is seen at the transition between triggers dominated by
  night sky background fluctuations, and those dominated by cosmic ray
  air showers. The dashed line indicates a typical threshold setting
  for standard operations. \newline {\bf Right:} Data products for a VERITAS
  telescope, consisting of a digitized signal trace for each PMT
  in the 499 pixel camera. The image in the camera has been cleaned,
  by setting the signal to zero in all pixels which contain no
  Cherenkov light. The ellipse shows a simple moment-based
  parameterization of the image. }
\label{bias}
\end{figure}

 Trigger systems typically work on multiple levels - the design goal
 being to trigger on the faintest possible Cherenkov flashes, without
 incurring a prohibitively high rate of false triggers due to the
 fluctuating night sky background. Individual pixels are equipped with
 discriminators, which produce a digital output. The outputs for each
 camera are passed to a logic circuit which looks for spatial
 coincidences (e.g. 3 neighbouring pixels must have triggered within a
 few nanoseconds). The final trigger decision occurs at the array
 level - typically at least 2 telescope cameras must have triggered at
 the same time, after correction for the different path lengths of the
 Cherenkov light to each telescope. Many variations on this basic
 scheme exist, notably the analog sum-trigger developed for use on the
 MAGIC telescopes.\cite{2014arXiv1404.4219G} Figure~\ref{bias} shows
 a``bias curve'' for the VERITAS array, illustrating the changing
 event rate as a function of individual pixel discriminator threshold,
 for each of the four telescopes, and for the complete array.

Prior to digitization, the photo-detector outputs are pre-amplified, as
close to the sensor as possible. This boosts the signal strength
without significantly increasing the signal-to-noise ratio, and allows
PMT detectors (in particular) to run at lower gain - extending their
useful lifetime, and protecting them from damage due to bright DC
light sources, including stars in the field-of-view. Since the
Cherenkov flashes sit upon a continuous DC background of night sky
light, the signals are also AC-coupled at this point. Digitization is
accomplished in various different ways - historically, simple
integrating Analog-to-Digital Converters (ADCs) were used, while more
modern systems use flash-ADCs or custom-designed analog ring sampling
devices, operating at multi-GHz sampling speeds. The final data
products consist of a sampled (or integrated) signal trace for every
pixel for each triggered event (Fig.~\ref{bias}). The data rate of
modern ACTs is in excess of a few hundred recorded events per second,
and reaches a few KHz for the largest telescopes.

\subsection{Peripheral and Environmental Systems}

In addition to the telescopes themselves, a wide variety of peripheral
systems are usually deployed, associated with calibration and
monitoring tasks, both of the telescopes and of the atmosphere above
them. Telescope calibration requires nanosecond light pulsers, used to
flat-field the photodetector gains, and to measure their single photon
response. Atmospheric monitoring can be achieved with local weather
stations, and with LIDARs, optical telescopes, and infra-red
radiometers, which can reveal the presence of clouds by measuring the
radiative temperature of the night sky.

\subsection{Array Design}
To this point, we have focussed on the design of individual
telescopes; however, the use of multiple telescopes in concert
dramatically increases the sensitivity of the technique, along with
its angular and spectral resolution.  Numerous studies have been
performed on the optimum layout and spacing of ACT
arrays.\cite{2000APh....13..253H, 2009APh....32..221C,
  2013APh....43..171B} The conclusions can broadly be summarized as
follows: (i) more telescopes are better, with the array sensitivity
increasing roughly as the square root of the number of telescopes, and
(ii) the optimal spacing depends upon the energy range to be covered:
wider spacing provides best sensitivity for higher energies. One
further point to note is that the array performance changes when the
array becomes significantly larger than the Cherenkov light
pool.\cite{2008ICRC....3.1441F} In this regime the central region of
the Cherenkov light pool is always sampled by multiple telescopes
(unlike smaller arrays, where the shower core often lies outside of
the area enclosed by the array).  The result of this is appreciably
better sensitivity at low energies, for telescopes of relatively
modest size.

\section{ACT Data Analysis}

The analysis of ACT data is complex, and the details have comparable
impact on the sensitivity and performance of the array as do many
aspects of the hardware. To recap, the goal of the the analysis is to
identify the primary particle, and to reconstruct its arrival
direction and energy. This information is then used to assess the
statistical significance of any gamma-ray signal, to map its
distribution on the sky, and to reconstruct the gamma-ray flux and energy
spectrum. Many different analysis methods exist in the literature, and
the details vary between the different arrays. Here we describe the
most common techniques in broad detail, and conclude with a
brief summary of some of the more sophisticated methods in use.

\subsection{Flat-fielding and Image Cleaning}

The raw data products for ACTs consist of a digitally sampled signal
trace for each of the photosensors in the cameras, roughly centered on
the arrival time of the Cherenkov pulse (Fig.~\ref{bias}). The first
stage of the data processing consists of measuring and subtracting the
signal \textit{pedestal} value - the baseline value in the absence of
any Cherenkov photons. The next step is to identify those pixels which
contain a Cherenkov signal, above some pre-defined threshold. The
signals are then corrected for variations in the photodetector gain
values, measured using a calibration light pulse. The result of this
pre-processing is a cleaned, calibrated image, typically approximately
elliptical in shape (Fig.~\ref{bias}).

\subsection{Identification of the Primary}
Even for a moderately strong gamma-ray source, cosmic ray shower
images outnumber gamma-ray shower images by at least a factor of
$\sim10^5$. Effective separation of the gamma-ray events is therefore
crucial. In the case of a point source of gamma rays, by far the most
effective tool for discrimation is the arrival direction of the
primary - but many TeV sources have large angular extent, up to a few
degrees in diameter. Fortunately, significant differences in the
Cherenkov image morphology, originating in the differences in the air
shower development, make discrimination possible - despite the
relatively crude optics and camera pixellation of an ACT.\cite{1985ICRC....3..445H}

The cleaned images are parameterized by a simple moment analysis, in
which their $width$, $length$ and orientation are
calculated. Gamma-ray images are typically less wide, and shorter,
than cosmic ray images with similar Cherenkov intensity and impact
parameter. In the case of a single telescope, simply selecting images
with small $width$ and $length$ provides fair
discrimination.\cite{1991ICRC....1..464P} The power of this analysis
is dramatically increased, however, when multiple telescopes view the
same shower. In this case, the shower core location, and hence the
impact distance from each telescope ($R$), can be determined to within
an accuracy of $\sim10\U{m}$. The core location is reconstructed
geometrically; in the reference frame of the array, all images point
away from the shower core location, and so the core can be found by
intersecting the image major axes.

Once the core location is known, the measured $width$ of the image can
be compared with a prediction, $width_{MC}$, for images with the same
Cherenkov intensity, $s$. This prediction, with an associated spread,
$\sigma_{width}$, is derived from detailed Monte Carlo simulations of
the air shower development and the telescope response. The predicted
widths are typically stored in look-up tables, a number of which are
generated corresponding to various different conditions under which
the observations were taken (e.g. elevation angle, background
night-sky brightness). The result of this comparison is then combined
for all of the Cherenkov images of the shower ($N_{images}$) like so:

\[
mscw=\frac{1}{N_{images}}\left[\sum_{i=1}^{N_{images}}\frac{width_i-width_{MC}(R,s)}{\sigma_{width}(R,s)}\right]
\]

$mscw$ is known as the``mean-scaled width'', and is used to provide
effective discrimation between gamma-ray and cosmic ray initiated
events\cite{1997APh.....8....1D, 1999APh....10..275H}
(Figure~\ref{MSCWnSkymap}). A similar method can be applied to the
image $length$. Various other parameters have also been derived and
used with different degrees of success (e.g. height of shower maximum,
Cherenkov photon arrival time gradient along the shower).

For the purposes of gamma-ray astronomy, a simple discrimination
between gamma rays and all other primaries is usually all that is
required. ACTs can also serve as powerful tools for cosmic ray
physics, however, and attempts have been made to measure the spectrum
and composition of the nuclear cosmic ray flux\cite{1999PhRvD..59i2003A, 2007PhRvD..75d2004A}, as well as the
electron component.\cite{2009AnA...508..561A}

\subsection{Arrival Direction Reconstruction}

Reconstruction of the arrival direction of the shower primary serves
two purposes: it provides effective discrimination between gamma-ray
photons from the source and the isotropic charged cosmic ray
background, and it allows us to study and to map out the gamma-ray
emission. Accurate location of the point of origin of the gamma-ray
emission is often necessary to the identification of gamma-ray
sources, and gamma-ray mapping of extended astrophysical sources
provides clues to the particle acceleration processes at work in these
objects.

In the field-of-view of the telescopes, the major axes of the image
ellipses intersect at the point corresponding to the arrival direction
of the primary particle, as shown schematically in
Figure~\ref{schematic}. This fact is used to provide an estimate of
the arrival direction, usually with some weighting scheme which gives
additional weight to the axes of the brightest images.\cite{1999APh....12..135H} The resulting angular resolution of the
technique is energy dependent, with typically $68\%$ of the gamma rays
from a point source reconstructed to within $0.1^{\circ}$ of the
source location, for energies around $1\U{TeV}$. At lower energies,
fluctuations in the shower development, and low Cherenkov photon
statistics, degrade the resolution somewhat.

Once the arrival directions have been calculated, any point in the
field-of-view can be tested for evidence for gamma-ray emission, by
selecting those events which lie within a pre-defined radius around
the test point. This process is complicated by the fact that the
gamma-ray emission from each point lies on top of the remaining
background of misidentified cosmic ray events. In order to calculate
the gamma-ray excess, and to calculate the statistical significance of
this excess, it is therefore necessary to find an independent estimate
of the remaining background at each point. This is accomplished by
measuring the background rate in blank regions of the sky, from which
little or no gamma-ray emission is expected.  These ``OFF-source''
regions can be selected in a variety of different ways: for example by
dedicated observations of adjacent fields-of-view, or (more commonly)
by selecting regions within the same field-of-view, but offset from
the test position. In this latter case, particular care must be taken
to account for the varying detection efficiency across the
field-of-view. A full description of a sample of common background
estimation techniques is given by Berge \textit{et al}\cite{2007AnA...466.1219B}.

\begin{figure}
\centerline{\includegraphics[width=1.0\textwidth]{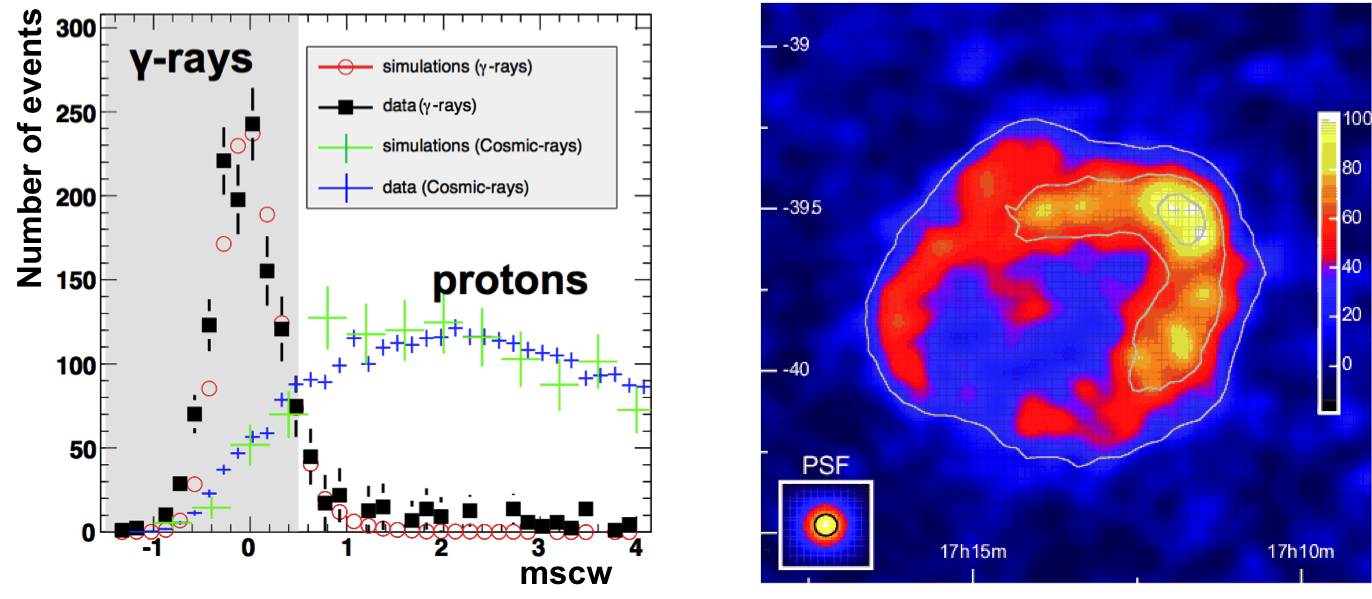}}
\caption{{\bf Left:} The distribution of the mean-scaled width for
  gamma-rays and cosmic rays, from Monte Carlo simulations and for a
  real gamma-ray source. Selecting events
  within the shaded region provides effective discrimination of
  gamma-ray events from the hadronic background \newline {\bf
    Right:} H.E.S.S. map of gamma-ray emission from the supernova
  remnant RX~J1713.7-3946.\cite{2007AnA...464..235A}}
\label{MSCWnSkymap}
\end{figure}

Once the background is known, the gamma-ray excess at any position can
be measured, and its significance calculated\cite{1983ApJ...272..317L} (or an
upper limit to the number of excess events, in the case of no
detection). By testing a range of positions on a 2-D grid, a map of
gamma-ray emission on the sky can be
constructed. Figure~\ref{MSCWnSkymap} shows an example of a gamma-ray
excess map from the direction of a supernova remnant, as measured by
the H.E.S.S. telescope array. Converting the excess (or upper limit)
into a measurement of the photon flux from the source (in$\U{photons}\UU{cm}{-2}\UU{s}{-1}$), requires detailed modelling of
the energy-dependent effective area of the telescope array, as
described in the following section.

\subsection{Gamma-Ray Energy, Flux and Spectrum}

Calculation of the energy of an incident gamma-ray relies upon the
fact that, to a good approximation, the number of particles in the
shower, and hence the Cherenkov photon yield, is directly proportional
to the primary energy. Measuring the Cherenkov emission intensity, and
combining this with the distance to the shower, therefore allows an
estimate of the gamma-ray energy.\cite{1998APh.....9...15M} Multiple
telescopes improve this energy estimate dramatically, since they
provide multiple measurements of the shower light yield, and an
improved estimate of the shower core location.\cite{2004ApJ...614..897A}

In practice, the energy estimate is made by referring to look-up
tables which contain the predicted gamma-ray energy as a function of
impact parameter and Cherenkov intensity. The contents of the tables
are derived from Monte Carlo simulations of the shower development,
and of the telescope response. For the purposes of energy estimation,
the most important parameter of the telescope model is the single
photo-electron response of the photo-detectors and their read-out
electronics. The most important factor in simulating the Cherenkov
yield at the telescope mirrors is the Earth's atmosphere.  This is
much more difficult to monitor and account for, and hence introduces
systematic uncertainties at the level of at least $10\%$. Numerous
tables are generated, corresponding to different observing conditions
(elevation angle, background night-sky brightness, source offset in
the field-of-view). The energy resolution of the technique depends
upon the energy of the primary, reaching typically $\sim15\%$ above
$1\U{TeV}$, and degrading below this.

Converting the measured energy distribution of gamma-rays from a
source into a meaningful flux estimate, or energy spectrum, requires
knowledge of the effective area of the detector. In the case of ACTs,
the maximum effective area is determined by the size of the Cherenkov
light pool, rather than by the size of the telescopes or the area of
the array, and can reach $>10^5\UU{m}{2}$ at high energies. At lower
energies, the trigger efficiency of the array (and hence the effective
area) drops sharply, eventually reaching zero. The energy-dependent
effective area is calculated by simulating gamma-ray showers over a
wide range of impact parameters, and with an energy distribution
similar to a typical source (e.g. a power law with an index of
$-2.0$). The ratio of the number of triggered events to the number of
events simulated, multiplied by the area over which the events were
thrown, then gives the effective area. The effective area again
depends upon a wide variety of operating conditions (elevation, sky
brightness) and analysis parameters (gamma-ray selection cuts, exact
analysis method), which must be precisely matched between the data
analysis and the simulations.

The reconstructed energy distribution of gamma-ray events from a
source can then be divided by the energy-dependent effective area in
order to reconstruct the true energy spectrum of the
source. Systematic biases can arise due to the fact that the effective
area estimate depends upon the simulated gamma-ray spectrum (due to
the finite energy resolution of the instrument). This can be addressed
by recalculating the effective area using the fitted energy spectrum
iteratively, until the two converge. More sophisticated unfolding
methods are also used to account for the finite resolution of the
technique.\cite{2007NIMPA.583..494A}

Gamma-ray source spectra are smooth continua, typically well-fit by
straight or curved power laws, or by a power-law with an exponential
cutoff. The spectra can be most easily parameterized by fitting a
chosen form to the gamma-ray flux points using the least-squares
method. A more sophisticated approach, less prone to biases introduced
by binning the data, is to perform a maximum-likelihood estimation of
the spectral parameters, taking into account the effective area and
the energy-resolution function of the detector.\cite{1999AnA...350...17D}

\subsection{Alternative Analysis Methods}

The analysis methods described above were developed by the Whipple and
HEGRA collaborations in the 1990s. They are robust against changing
conditions, provide good sensitivity, and are widely used to this
day. However, the development of analysis tools has always proceeded
in parallel with the hardware developments of ACTs, and many
alternative methods exist in the literature. Some of these provide
significance improvements in sensitivity, energy threshold, or angular
or spectral resolution.

One flaw of the standard method is that it does not take advantage of
the fact that an array provides multiple views of the same shower, and
so the images recorded should be correlated. This additional
information can be exploited by performing a global fit to the data,
using a model of the shower development based on the primary energy,
arrival direction and impact parameter. The first implementation of
this method was made by the CAT collaboration, using just a single
telescope, and a simple analytical 2-D model of the shower
profile.\cite{1998NIMPA.416..425L} The technique has subsequently been
refined to work with multiple telescopes, to perform a log-likelihood
minimisation using all pixels in the camera\cite{2009APh....32..231D},
and to use 3D analytical models of the shower
development\cite{2006APh....25..195L}, or direct comparison with
template images generated by Monte Carlo
simulations.\cite{2014APh....56...26P} In these schemes, the
goodness-of-fit parameter provides a single powerful discriminant to
separate gamma-rays from background. The method also automatically
provides an energy estimate, which can be used to reconstruct spectra.

Another approach is to improve the discrimination between gamma-ray
and cosmic ray events through the use of advanced pattern recognition
or multivariate analysis techniques. Some of the most successful
approaches to this draw on developments in the field of experimental
particle physics, where similar problems of classification are often
met. While many different techniques have been attempted (neural
networks, genetic algorithms, etc.), the most efficient appear to be
the decision tree methods: boosted decision trees\cite{2009APh....31..383O,2010APh....34...25F,2011APh....34..858B} and
random forests.\cite{2008NIMPA.588..424A} Inputs to these machine
learning algorithms can correspond to the simple geometrical
parameters of the standard analysis method, or encompass additional
information, including the results of the template fitting methods
described above.

Finally, many attempts have been made to explore additional properties
of the Cherenkov radiation from air showers, in the hope of finding
complementary information to enhance the analysis. Some have failed -
the spectrum\cite{1996SSRv...75...17H} or the polarization\cite{2002APh....17..133D} of the Cherenkov light, for example, do not
seem likely to provide any useful additional discrimination. The
arrival time of Cherenkov photons, however, does improve
discrimination somewhat - an aspect of the analysis that becomes more
important with the development of very large isochronous reflectors,
and very fast ($\geq1\U{GHz}$) sampling electronics.\cite{1999APh....11..363H, 2009APh....30..293A, 2011APh....34..886S}

\section{Concluding Remarks}
Atmospheric Cherenkov gamma-ray telescopes have proven remarkably
successful over the past decade. Small arrays of moderately-sized
telescopes have opened a new window on the Universe, probing particle
acceleration in extreme environments both within and outside of our
Galaxy. The next stage in the development of the technique requires
substantial investment, and hence collaboration on a global
scale. This is proceeding through the ``Cherenkov Telescope Array''
(CTA) project, which is designing and constructing a next generation
instrument.\cite{2013APh....43....3A} The plan involves a km$^2$
array with a few large aperture ($\sim23\U{m}$) telescopes at the
centre, surrounded by an array of moderately-sized telescopes
($\sim10\U{m}$) with $\sim100\U{m}$ spacing, supplemented by a
wider-spaced array of smaller telescopes ($4\U{m}$). A graded array
such as this is expected to provide sensitivity improvements of an
order of magnitude over current arrays, together with the widest
possible energy coverage. Prototyping and testing is underway, and new
technologies are being tested at all stages (e.g. in mirror designs,
photosensors, and trigger and data acquisition systems). The goal of
such development is not only to enhance the array performance, but
also to deal with the necessities of mass production, low cost, and
strict maintenance requirements.  Both northern and southern
hemisphere arrays are envisaged, and possible sites are currently
under discussion.

\bibliographystyle{ws-rv-van}
\bibliography{jholder_ACTs_bib}
\blankpage
\printindex                         
\end{document}